\documentclass[conference]{IEEEtran}
\IEEEoverridecommandlockouts
\usepackage{cite}
\usepackage{amsmath,amssymb,amsfonts}
\usepackage{multirow}
\usepackage{mathtools}
\usepackage{amsthm}
\theoremstyle{definition}
\newtheorem{assumption}{Assumption}
\newtheorem{remark}{Remark}
\newtheorem{theorem}{Theorem}
\newtheorem{lemma}{Lemma}
\newtheorem{definition}{Definition}

\usepackage{algorithmic}
\usepackage{graphicx}
\usepackage{textcomp}
\usepackage{xcolor}
\usepackage{comment}
\def\BibTeX{{\rm B\kern-.05em{\sc i\kern-.025em b}\kern-.08em
    T\kern-.1667em\lower.7ex\hbox{E}\kern-.125emX}}
\begin{document}

\title{Decentralized Hidden Markov Modeling with Equal Exit Probabilities
\thanks{This work was conducted while D.S. was a visiting PhD student at Imperial College London with support from the China Scholarship Council (No. 202406100217). S.L. was supported by the National Natural Science Foundation of China (No. 12471457).}
}

\author{\IEEEauthorblockN{Dongyan Sui}
\IEEEauthorblockA{\textit{Fudan University}\\
Shanghai, China \\
dysui22@m.fudan.edu.cn}
\and
\IEEEauthorblockN{Haitian Zheng}
\IEEEauthorblockA{\textit{Fudan University}\\
Shanghai, China \\
htzheng24@m.fudan.edu.cn}
\and
\IEEEauthorblockN{Siyang Leng}
\IEEEauthorblockA{\textit{Fudan University}\\
Shanghai, China \\
syleng@fudan.edu.cn}
\and
\IEEEauthorblockN{Stefan Vlaski}
\IEEEauthorblockA{\textit{Imperial College London}\\
London, United Kingdom \\
s.vlaski@imperial.ac.uk}
}

\maketitle

\begin{abstract}
Social learning strategies enable agents to infer the underlying true state of nature in a distributed manner by receiving private environmental signals and exchanging beliefs with their neighbors. Previous studies have extensively focused on static environments, where the underlying true state remains unchanged over time. In this paper, we consider a dynamic setting where the true state evolves according to a Markov chain with equal exit probabilities. Based on this assumption, we present a social learning strategy for dynamic environments, termed Diffusion $\alpha$-HMM. By leveraging a simplified parameterization, we derive a nonlinear dynamical system that governs the evolution of the log-belief ratio over time. This formulation further reveals the relationship between the linearized form of Diffusion $\alpha$-HMM and Adaptive Social Learning, a well-established social learning strategy for dynamic environments. Furthermore, we analyze the convergence and fixed-point properties of a reference system, providing theoretical guarantees on the learning performance of the proposed algorithm in dynamic settings. Numerical experiments compare various distributed social learning strategies across different dynamic environments, demonstrating the impact of nonlinearity and parameterization on learning performance in a range of dynamic scenarios.
\end{abstract}


\section{Introduction and Related Work}
Social learning refers to the process by which networked agents infer the underlying true state of the environment by gathering information and sharing beliefs. In multi-agent or social networks, Bayesian and non-Bayesian social learning models \cite{acemoglu2011bayesian,jadbabaie2012non,molavi2018theory,nedic2017fast,lalitha2018social} have been extensively employed in economics, sociology, and engineering to characterize the behaviors of financial markets, social groups, and multi-agent systems\cite{chamley2004rational,acemoglu2011opinion,bordignon2022learning}. Traditional models have mainly focused on static environments. Recent research has increasingly explored online social learning models in dynamic settings, such as adaptive social learning\cite{9475073,10715423} and diffusion hidden Markov modeling strategies\cite{kayaalp2022hidden,kayaalp2022distributed}. 

In this paper, we consider the online social learning problem, where a network of $N$ agents labeled by $k=1,\ldots, N$ receive noisy observations/signals $\boldsymbol{\xi}_{k,i}$ (bold notation is used for random variables) of the evolving state at each time step $i\ge1$. Their aim is to collectively estimate the underlying true state $\boldsymbol{\theta}_i^{\star}$ at each time instant $i$ given the streaming observations $\boldsymbol{\xi}_{k,1},\ldots,\boldsymbol{\xi}_{k,i}$.

For simplicity, we assume that the true state $\boldsymbol{\theta}_i^{\star}$ belongs to a discrete set of $M$ possible states $\Theta = \{\theta_0, \theta_1, \ldots, \theta_{M-1}\}$. Each agent $k$ assigns a belief to each state $\theta\in\Theta$ at each time step $i$, denoted by $\boldsymbol{\mu}_{k,i}(\theta)$. The belief characterizes the agents' confidence that $\theta$ is the underlying true state at time $i$ and is a probability distribution over all possible states $\Theta$, i.e., $\sum_{m=0}^{M-1}\boldsymbol{\mu}_{k,i}(\theta_m)=1$, for all $i=0,1,\ldots$ and $k=1,\ldots,N$. To avoid triviality, we assume that each agent's initial belief, $\boldsymbol{\mu}_{k,0}(\theta)$, is strictly positive for all $\theta \in \Theta$. Correct learning is said to occur at time $i$ for agent $k$ if the belief $\boldsymbol{\mu}_{k,i}(\theta)$ is maximized at the true state $\theta = \boldsymbol{\theta}_i^{\star}$. The observations $\boldsymbol{\xi}_{k,i}$ are independent random variables over time $i$ conditioned on the true state \( \boldsymbol{\theta}_i^{\star} \), taking values in the space $\Xi_k$. Given the underlying true state $\theta_i^{\star}$, the observations follow a probability density function $f(\cdot|\theta_i^{\star})$, which implies that when the underlying state of the environment remains unchanged, the observations are independent and identically distributed (i.i.d.) random variables over time. Each agent $k$ is equipped with a model that specifies the likelihood of the observations $\xi\in\Xi_k$ for each possible state $\theta \in \Theta$, denoted by $L_k(\xi|\theta)$.

The likelihood model $L_k(\xi|\theta)$, as a function of $\xi$, can be either a probability density function or a probability mass function, depending on whether $\xi$ is continuous or discrete. To ensure the agents can successfully learn the underlying true state, we impose the following assumptions, which are also the typical assumptions in traditional social learning methods\cite{jadbabaie2012non,nedic2017fast,lalitha2018social,9475073,10714918}:
\begin{assumption}[Finiteness of KL Divergence]
For each pair of distinct states $\theta$ and $\theta'$, the Kullback–Leibler (KL) divergence\cite{thomas2006elements} between $L_k(\xi \mid \theta)$ and $L_k(\xi \mid \theta')$ for any agent $k$ satisfies $D_{\text{KL}}(L_k(\xi|\theta)||L_k(\xi|\theta'))<\infty$.
\end{assumption}

This assumption avoids trivial cases where a likelihood function model for a certain state completely dominates.

\begin{assumption}[Global Identifiability of the Underlying True State]
The agents are collectively able to identify the true state uniquely:
\begin{equation}
\left\{\theta_i^{\star}\right\}=\Theta_i^{\star}=\bigcap_{k=1}^N\Theta_{k,i}^{\star},
\end{equation}
where $\Theta_{k,i}^{\star}=\underset{\theta\in\Theta}{\arg\min}D_{\text{KL}}(f(\cdot|\theta_i^{\star})||L_k(\cdot|\theta))$.
\end{assumption}
The agents interact in a network. We denote by $A=\left[a_{\ell k}\right]$ the weight matrix of the network, which is assumed to be \textit{nonnegative and left-stochastic}, i.e.,
$0\le a_{\ell k}\le 1$, $\sum_{\ell=1}^Na_{\ell k}=1$, $a_{\ell k}=0\text{ for } \ell\notin \mathcal{N}_k$, where $\mathcal{N}_k$ denotes the neighborhood of agent $k$, with $k$ itself being included.

\begin{assumption}[Strong connectivity and aperiodicity]
The network of agents is strongly connected, and at least one node $k$ in the network has a self-loop, i.e., $a_{kk} > 0$.
\end{assumption}

Under Assumption 3, the weight matrix $A$ is a primitive matrix. According to the Perron-Frobenius theorem \cite{horn2012matrix}, there exists a Perron vector $\pi$ satisfying:
\begin{equation}\label{perron}
A\pi=\pi,\quad\sum_{k=1}^N\pi_k=1,\quad\pi_k>0\text{ for all }k=1,\ldots,N.
\end{equation}

\subsection{Diffusion $\alpha$-HMM}
We model the true state \( \boldsymbol{\theta}_{i}^{\star} \) as a random variable, following a Markov chain. If the transition probabilities are denoted by $P = \left[p_{nm}\right]_{M \times M}$, where $p_{nm} = \mathbb{P}\left[\theta_i^{\star} = \theta_m \mid \theta_{i-1}^{\star} = \theta_n \right]$, then the optimal private belief update, based on the hidden Markov model (HMM), is given by:
\begin{equation}\label{hmm}
\boldsymbol{\psi}_{k,i}(\theta_m)=\frac{\sum_{n=0}^{M-1}p_{nm}\boldsymbol{\mu}_{k,i-1}(\theta_n)L_k(\boldsymbol{\xi}_{k,i}|\theta_m)}{\sum_{\ell=0}^{M-1}\sum_{n=0}^{M-1}p_{n\ell}\boldsymbol{\mu}_{k,i-1}(\theta_n)L_k(\boldsymbol{\xi}_{k,i}|\theta_{\ell})}.
\end{equation}

This type of private belief update rule in social learning has been studied in \cite{kayaalp2022hidden,kayaalp2022distributed}. Observe that the optimal private belief update under a hidden Markov model involves the entries of the full state transition matrix $P$, which is frequently unknown and challenging to estimate in practice. In this work, we will instead study a simplified HMM-based update, which is derived under the assumption of equal exit probabilities for state transitions. Specifically, the true state transitions with a probability $h$, and when a transition occurs, the next state is chosen uniformly at random. Under these assumptions, the transition probability matrix simplifies to $p_{mn}=1-h$ if $m=n$, and $p_{mm}=h/(M-1)$.

A similar transition model was used in~\cite{10715423} to quantify the dynamics of Adaptive Social Learning in time-varying environments. This transforms the traditional HMM belief update into:
\begin{equation}\label{inference_alpha_hmm}
\boldsymbol{\psi}_{k,i}(\theta_m)=\frac{((1-\alpha M)\boldsymbol{\mu}_{k,i-1}(\theta_m)+\alpha)L_k(\boldsymbol{\xi}_{k,i}|\theta_m)}{\sum_{n=0}^{M-1}((1-\alpha M)\boldsymbol{\mu}_{k,i-1}(\theta_n)+\alpha)L_k(\boldsymbol{\xi}_{k,i}|\theta_n)},
\end{equation}
where $\alpha=\frac{h}{M-1}$ represents the exit probability. This private belief update rule, referred to as the $\alpha$-HMM, simplifies the inference problem to a single tunable hyperparameter $\alpha$. Of course, for a general state transition matrix $P$, the simplified update rule \eqref{inference_alpha_hmm} will be suboptimal compared to the optimal HMM filter \eqref{hmm}. The advantage, on the other hand, is that this simplification enhances the theoretical tractability, and \eqref{inference_alpha_hmm} relies only on a single parameter $\alpha$, which quantifies the volatility of the underlying true state. Indeed, examining \eqref{inference_alpha_hmm}, we observe that $\alpha$ essentially controls the amount of weight placed on prior beliefs compared to the most recent observation. Note that when $\alpha=0$, the above iteration degenerates into the classical Bayes' update:
\begin{equation}\label{Bayes}
\boldsymbol{\psi}_{k,i}(\theta_m)=\frac{\boldsymbol{\mu}_{k,i-1}(\theta_m)L_k(\boldsymbol{\xi}_{k,i}|\theta_m)}{\sum_{n=0}^{M-1}\boldsymbol{\mu}_{k,i-1}(\theta_n)L_k(\boldsymbol{\xi}_{k,i}|\theta_n)}.
\end{equation}
On the other hand, when $\alpha = 1/M$, Eq.~\eqref{inference_alpha_hmm} reduces to:
\begin{equation}\label{alpha1M}
\boldsymbol{\psi}_{k,i}(\theta_m) = \frac{L_k(\boldsymbol{\xi}_{k,i} | \theta_m)}{\sum_{n=0}^{M-1} L_k(\boldsymbol{\xi}_{k,i} | \theta_n)}.
\end{equation}
In this case, each agent will negate prior beliefs and rely solely on the current private observation for learning. In this paper, since we focus on filtering in dynamic environments, we only consider the case where $0<h<1$ and consequently, $\alpha>0$.

At time $i$, after the inference step~\eqref{inference_alpha_hmm}, each agent $k$ aggregates private beliefs from its neighboring nodes to form its current belief $\boldsymbol{\mu}_{k,i}$ using a geometrically weighted average:
\begin{equation}\label{aggregation_alpha_hmm}
\boldsymbol{\mu}_{k,i}(\theta_m)=\frac{\exp\left(\sum_{\ell\in\mathcal{N}_k}a_{\ell k}\log\boldsymbol{\psi}_{\ell,i}(\theta_m)\right)}{\sum_{n=0}^{M-1}\exp\left(\sum_{\ell\in\mathcal{N}_k}a_{\ell k}\log\boldsymbol{\psi}_{\ell,i}(\theta_n)\right)}.
\end{equation}

Combining \eqref{inference_alpha_hmm} and \eqref{aggregation_alpha_hmm} together, we finally obtain the \textit{Diffusion $\alpha$-HMM strategy} for social learning in dynamical environment. Note that the strategy corresponds to a simplified form of the Diffusion HMM strategy from~\cite{kayaalp2022distributed, kayaalp2022hidden}. In contrast to these works, we will not assume the Markov chain to be consistent with the dynamics driving \( \boldsymbol{\theta}_i^{\star}\), and instead treat \( \alpha \) as a tunable parameter akin to a step-size. 

\subsection{Dynamics of Log-belief Ratio in Steady State}
For the purposes of analysis, we assume the environment remains in a single state over an extended period, resulting in fixed observation statistics. Without loss of generality, we assume that the underlying true state is $\theta_0 \in \Theta$, i.e., $\theta_i^{\star}=\theta_0$ for all $i\ge1$. We aim to analyze the dynamics of the log-likelihood ratio of the belief on wrong and true states for any agent $k$, i.e.:
\begin{equation}
\boldsymbol{x}_{k,i}(\theta_m)\triangleq \log\frac{\boldsymbol{\mu}_{k,i}(\theta_m)}{\boldsymbol{\mu}_{k,i}(\theta_0)},\quad m=1,\ldots,M-1.
\end{equation}

It can be verified that the log-likelihood ratio $\boldsymbol{x}_{k,i}(\theta_m)$ in Diffusion $\alpha$-HMM evolves as:
\begin{equation}\label{log-belief-ratio-dynamics}
\boldsymbol{x}_{k,i}(\theta_m)=\sum_{\ell\in\mathcal{N}_k}a_{\ell k}\left(F_m(\boldsymbol{x}_{\ell,i-1})+\log\frac{L_{\ell}(\boldsymbol{\xi}_{\ell,i}|\theta_m)}{L_{\ell}(\boldsymbol{\xi}_{\ell,i}|\theta_0)}\right),
\end{equation}
where
\begin{equation}\label{Fm}
\begin{aligned}
&F_{m}(x_1,\ldots,x_{M-1})\triangleq\\
&\log\frac{(1-\alpha M)\exp(x_{m})+\alpha+\alpha\sum_{n=1}^{M-1}\exp(x_{n})}{1-\alpha M+\alpha+\alpha\sum_{n=1}^{M-1}\exp(x_{n})},
\end{aligned}
\end{equation}
\begin{equation}
\boldsymbol{x}_{k,i}=\left[\boldsymbol{x}_{k,i}(\theta_1),\ldots,\boldsymbol{x}_{k,i}(\theta_{M-1})\right]^\top.
\end{equation}

The following remark shows the connection between Diffusion $\alpha$-HMM and the Adaptive Social Learning strategy from~\cite{9475073}.
\begin{remark}
The nonlinear function \eqref{Fm} has the following properties:
\begin{equation}
F_m(0,\ldots,0)=0,
\end{equation}
\begin{equation}
\frac{\partial F_m}{\partial x_m}\Big|_{(0,\ldots,0)}=1-\alpha M,
\end{equation}
\begin{equation}
\frac{\partial F_m}{\partial x_n}\Big|_{(0,\ldots,0)}=0,\quad \forall n\neq m.
\end{equation}
By applying a multivariate Taylor expansion to $F(x)$ around $x = 0_{M-1}$ up to the first-order term, for all $m=1,\ldots,M-1$ we have $F_m(x_1,\ldots,x_{M-1})=(1-\alpha M)x_{m}+o(\left\|x\right\|)$. Then, the linear approximation of the system \eqref{log-belief-ratio-dynamics} is given by:
\begin{equation}\label{linearized_ahmm}
\begin{aligned}
\boldsymbol{x}_{k,i}(\theta_m)=&(1-\alpha M)\sum_{\ell\in\mathcal{N}_k}a_{\ell k}\boldsymbol{x}_{\ell,i-1}(\theta_m)\\
&+\sum_{\ell\in\mathcal{N}_k}a_{\ell k}\log\frac{L_{\ell}(\boldsymbol{\xi}_{\ell,i}|\theta_m)}{L_{\ell}(\boldsymbol{\xi}_{\ell,i}|\theta_0)}.
\end{aligned}
\end{equation}
The above equation resembles the evolution of the log-belief ratio in the Adaptive Social Learning (ASL) strategy \cite{9475073}, which has the following form:
\begin{equation}\label{asl}
\begin{aligned}
\boldsymbol{x}_{k,i}(\theta_m)=&(1-\delta)\sum_{\ell\in\mathcal{N}_k}a_{\ell k}\boldsymbol{x}_{\ell,i-1}(\theta_m)\\
&+\delta\sum_{\ell\in\mathcal{N}_k}a_{\ell k}\log\frac{L_{\ell}(\boldsymbol{\xi}_{\ell,i}|\theta_m)}{L_{\ell}(\boldsymbol{\xi}_{\ell,i}|\theta_0)},
\end{aligned}
\end{equation}
where $\delta$ is the step-size parameter. Equations \eqref{linearized_ahmm} and \eqref{asl} both apply a discount to the information from the previous time step in the private belief update step. The subtle difference between them lies in the fact that the latter normalizes the weighting between past and new information through the step-size parameter $\delta$. Further comparison of the performance among the Diffusion $\alpha$-HMM, linearized Diffusion $\alpha$-HMM, and ASL will be illustrated in numerical experiments.
\end{remark}

It can be seen that the recursion \eqref{log-belief-ratio-dynamics} is both \emph{nonlinear} and \emph{stochastic}. To facilitate analysis, we introduce the following deterministic dynamical reference system, where stochastic quantities are replaced by their expected values:
\begin{equation}\label{reference_dynamics}
\hat{x}_{k,i}(\theta_m)=\sum_{\ell\in\mathcal{N}_k}a_{\ell k}\left(F_m(\hat{x}_{\ell,i-1})-d_{\ell}(\theta_m)\right),
\end{equation}
where
\begin{align}
d_k(\theta_m)\triangleq&D_{\text{KL}}(f(\cdot|\theta_0)||L_k(\cdot|\theta_m))-D_{\text{KL}}(f(\cdot|\theta_0)||L_k(\cdot|\theta_0))\notag\\
=&-\mathbb{E}\left[\log\frac{L_k(\boldsymbol{\xi}_{k,i}|\theta_m)}{L_k(\boldsymbol{\xi}_{k,i}|\theta_0)}\right].
\end{align}
Assumption 2 ensures that $d_k(\theta_m)\ge 0$ and for all $m=1,\ldots,M-1$ there exists at least one $k=1,\ldots,N$ such that $d_k(\theta_m)>0$. The quantity $d_k(\theta_m)$ quantifies agent $k$'s ability to distinguish between an incorrect state $\theta_m$ and the true state $\theta_0$. We refer to this measure as the identifiability of agent $k$ with respect to $\theta_m$.

In this paper, we first analyze the fixed point of the dynamical reference system \eqref{reference_dynamics} under steady-state conditions. We then establish the convergence of the reference system to its fixed point. Finally, under additional assumptions on noise and identifiability, we derive an estimate for the error probability.

\section{Convergence Analysis}
For a discrete dynamical system $x_i = T(x_{i-1})$, a fixed point corresponds to an equilibrium state of the system, such that \( T(x^{\infty}) = x^{\infty} \). We begin by characterizing \( x_k^{\infty} \) for the reference system~\eqref{reference_dynamics}.

\begin{lemma}
The fixed points of the dynamical reference system \eqref{reference_dynamics} exist, and satisfy $\hat{x}^\infty_{k}(\theta_m)<-\sum_{\ell\in\mathcal{N}_k}a_{\ell k}d_{\ell}(\theta_m)$.
\end{lemma}
\begin{proof}
Proof omitted due to space limitations.
\end{proof}
The following theorem shows the convergence of \eqref{reference_dynamics} to its unique fixed point provided that the underlying state is constant.
\begin{theorem}
When $0<\alpha<1/M$ and \( \boldsymbol{\theta}_i^{\star} \) remains constant, $\hat{x}_{k,i}(\theta_m)$ defined in \eqref{reference_dynamics} will converge to a unique fixed point, i.e.,
\begin{equation}\label{lim_x}
\lim_{i\rightarrow\infty}\hat{x}_{k,i}(\theta_m)=\hat{x}_k^\infty(\theta_m),\quad \forall 1\le k\le N,\;1\le m\le M-1.
\end{equation}
\end{theorem}

\begin{proof}
Proof omitted due to space limitations.
\end{proof}

Theorem 1 establishes that the dynamical reference system converges to a unique fixed point, whose value is upper-bounded by a weighted average of the identifiability of neighboring agents, referred to as the neighborhood identifiability. Thus, neighborhood identifiability guarantees the learning capability of agent $k$ under steady-state conditions. 

In the following, for the purpose of further theoretical analysis, we make the following assumption, which requires the log-likelihood to be bounded almost surely.

%\begin{assumption}[Local Identifiability of the Underlying True State]
%For each agent $k$ and each pair of states $\theta$ and $\theta'$, there exists agent $l\in\mathcal{N}_k$ who can distinguish the two states based on its observations and likelihood model, i.e.,
%\begin{equation}
%\left\{\theta_i^{\star}\right\}=\bigcap_{\ell\in\mathcal{N}_k}^N\Theta_{\ell,i}^{\star},\quad\forall k=1,\ldots,N.
%\end{equation}
%where $\Theta_{\ell,i}^\star$ is as defined in %\eqref{identifiable}.
%\end{assumption}

%Assumption 4 guarantees that $\sum_{\ell\in\mathcal{N}_k}a_{\ell k}d_{\ell}(\theta_m)>0$ for all agent $k$ and state $\theta_m$.

\begin{assumption}[Bounded Log-Likelihood Ratio]
There exists a positive constant $C$ such that:
\begin{equation}\label{C_def}
\underset{k=1,\ldots,N}{\max}\:\underset{m=1,\ldots,M-1}{\max}\:\underset{\boldsymbol{\xi}\in\Xi_k}{\sup}\: \left|\log\frac{L_k(\boldsymbol{\xi}|\theta_m)}{L_k(\boldsymbol{\xi}|\theta_0)}+d_k(\theta_m)\right|\le C,\: a.s..
\end{equation}
\end{assumption}

To evaluate the learning performance of Diffusion $\alpha$-HMM, we introduce an important metric: the \textit{instantaneous error probability}, defined as:
\begin{equation}\label{error_prob}
\begin{aligned}
&p^e_{i}\triangleq \\ &\mathbb{P}\left[\;\exists\;k=1,\ldots,N\text{ and }\theta_m \neq \theta_0, \text{ s.t. } \boldsymbol{\mu}_{k,i}(\theta_m) \geq \boldsymbol{\mu}_{k,i}(\theta_0)\right].
\end{aligned}
\end{equation}
From this definition, it is clear that the instantaneous error probability quantifies the probability that any agent fails to correctly identify the underlying true state at a given time $i$ during the online social learning process.

Due to the nonlinearity and stochasticity of the original system, directly computing the error probability is  challenging. Here, we provide an analytical result under the assumption of sufficient neighborhood identifiability. 

\begin{theorem}
When $\alpha < 1/M$ and the underlying true state remains constant, i.e., $\theta_i^{\star}= \theta^{\star}$ for all $i=1,2,\ldots$, if for all $k=1,\ldots,N$ and $m=1,\ldots,M-1$,
\begin{equation}\label{data_quality}
\sum_{\ell\in\mathcal{N}_k}a_{\ell k}d_{\ell}(\theta_m)>C,
\end{equation}
then the instantaneous error probability for the diffusion $\alpha-$HMM algorithm satisfies:
\begin{equation}\label{ep_ineq}
\lim_{i\rightarrow \infty}p_i^e\le\frac{C}{-\alpha\overline{x}^\infty},
\end{equation}
where $\overline{x}^\infty=\underset{k=1,\ldots,N}{\max}\:\underset{m=1,\ldots,M-1}{\max}\hat{x}_k^\infty(\theta_m)<0$, $C$ is as defined in \eqref{C_def}.
\end{theorem}

\begin{proof}
Proof omitted due to space limitations.
\end{proof}

Theorem 2 establishes that, under certain conditions on neighborhood identifiability, the upper bound of the instantaneous error probability converges to a fixed value. This result implies that within the Diffusion $\alpha$-HMM framework, the probability of erroneous learning does not asymptotically approach zero, even in steady-state conditions, which contrasts with the behavior predicted by Bayes' formula \cite{lalitha2018social}. However, by sacrificing some learning accuracy, the Diffusion $\alpha$-HMM framework significantly enhances adaptability, as further demonstrated in the numerical experiments.

The steady-state error probability can be mitigated in one of two ways:  
(1) Shifting the fixed point $\hat{x}_k^\infty(\theta_m)$ of the deterministic system further away from zero. As shown in Lemma 1, improving agents' identifiability $d_k(\theta_m)$ facilitates this shift.  
(2) Reducing observation noise, thereby increasing the informativeness of the received signals.

\section{Numerical Experiments}

In the numerical experiments, we consider a network of $N=5$ agents attempting to infer the evolving state from a set of $M=3$ possible states. Different network topologies are examined, including a fully connected network and a strongly connected network. The weight matrix is randomly initialized while ensuring it satisfies the conditions of a primitive matrix. The true distribution of the observation $\boldsymbol{\xi}_{k,i}$, $f(\cdot|\theta_i^\star)$, follows a normal distribution $\mathcal{N}(\theta_i^\star,\sigma^2)$, where $\theta_i^\star \in \Theta = \{0,1,2\}$. Each agent $k$ employs a likelihood model $L_k(\cdot|\theta)$, which is also modeled as a normal probability density function with a standard deviation of $\sigma$ and a mean specified in Table~\ref{tab1}. As observed from Table~\ref{tab1}, some agents are unable to independently distinguish among the three states. However, for any pair of states, there exists at least one agent capable of differentiation, which satisfies Assumption~2.

\begin{table}[b]
\caption{Likelihood model configuration for the agents.}
\begin{center}
\begin{tabular}{|c|ccc|}
\hline
\multirow{2}{*}{\textbf{Agent $k$}} & \multicolumn{3}{c|}{\textbf{Likelihood model $L_k(\cdot|\theta)$}}                                      \\ \cline{2-4} 
                                      & \multicolumn{1}{c|}{\textit{\textbf{$\theta_0=0$}}} & \multicolumn{1}{c|}{\textit{\textbf{$\theta_1=1$}}} & \textit{\textbf{$\theta_2=2$}} \\ \hline
1                                     & \multicolumn{1}{c|}{0}                              & \multicolumn{1}{c|}{1}                              & 2                              \\ \hline
2                                     & \multicolumn{1}{c|}{0}                              & \multicolumn{1}{c|}{1}                              & 1                              \\ \hline
3                                     & \multicolumn{1}{c|}{0}                              & \multicolumn{1}{c|}{0}                              & 2                              \\ \hline
4                                     & \multicolumn{1}{c|}{0}                              & \multicolumn{1}{c|}{1}                              & 0                              \\ \hline
5                                     & \multicolumn{1}{c|}{0}                              & \multicolumn{1}{c|}{0}                              & 0                              \\ \hline
\end{tabular}
\label{tab1}
\end{center}
\end{table}

We compare the performance of three algorithms: the Diffusion $\alpha$-HMM introduced in this paper (Eq.~\eqref{inference_alpha_hmm} and Eq.~\eqref{aggregation_alpha_hmm}), the linearized Diffusion $\alpha$-HMM~\eqref{linearized_ahmm}, and adaptive social learning~\cite{bordignon2022learning}. Specifically, consider the private belief update given by:
\begin{equation}
\boldsymbol{\psi}_{k,i}(\theta_m)=\frac{\boldsymbol{\mu}^{1-\delta_1}_{k,i-1}(\theta_m)L^{\delta_2}_k(\boldsymbol{\xi}_{k,i}|\theta_m)}{\sum_{n=0}^{M-1}\boldsymbol{\mu}^{1-\delta_1}_{k,i-1}(\theta_n)L^{\delta_2}_k(\boldsymbol{\xi}_{k,i}|\theta_n)}.
\end{equation}
When $\delta_1 = \delta_2 = \delta$, this update rule corresponds to the adaptive social learning (ASL) strategy. When $\delta_1 = \alpha M$ and $\delta_2 = 0$, it represents the linearized Diffusion $\alpha$-HMM approach.

\begin{figure}[!t]
\centerline{\includegraphics[width=\columnwidth]{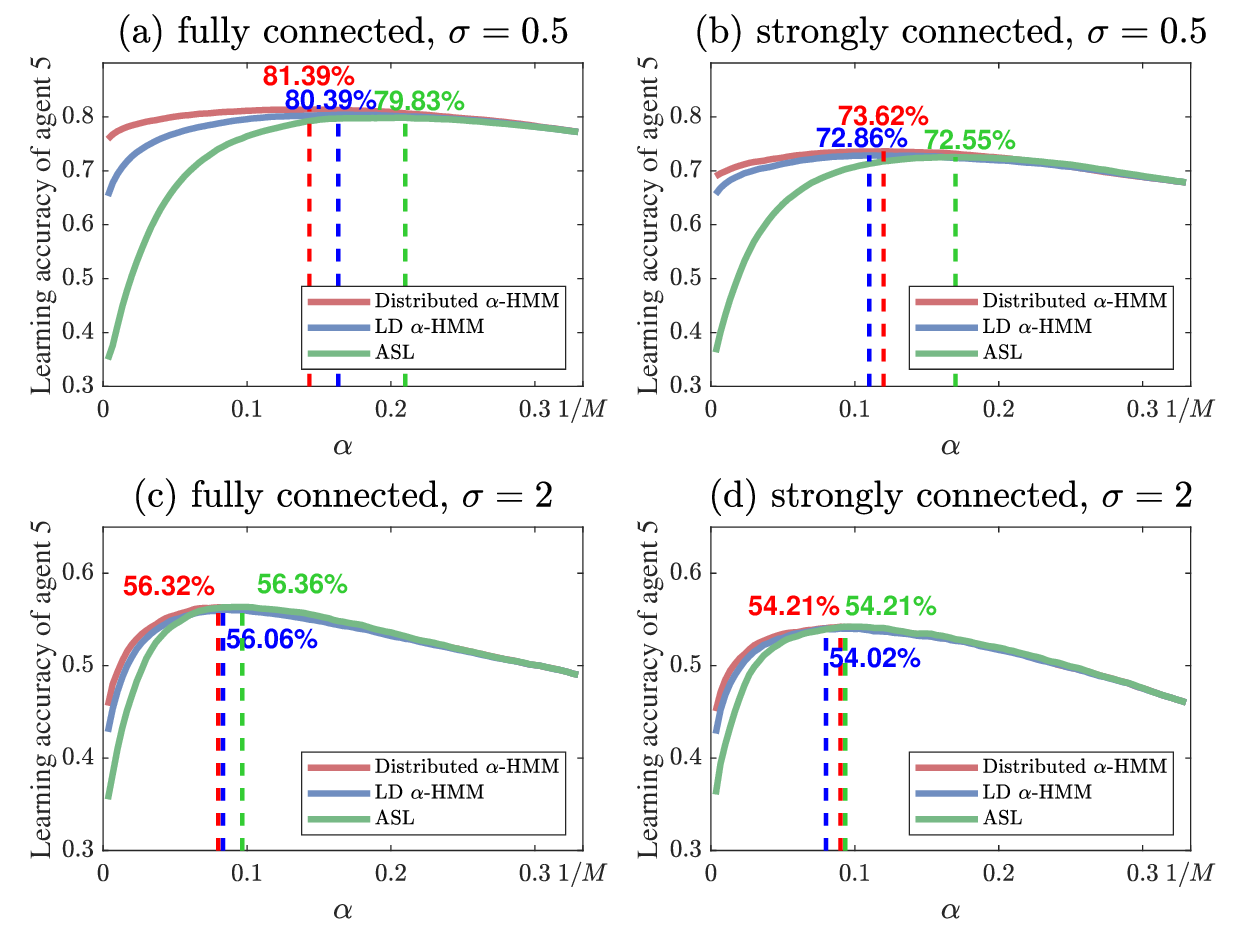}}
\caption{Comparison of the learning accuracy of agent 5, which has no distinguishing capability, under different network topologies and algorithms in Scenario 1. The accuracy is evaluated as a function of $\alpha$ for different fixed values of $\sigma$. LD $\alpha$-HMM refers to the linearized Diffusion $\alpha$-HMM.}
\label{NE_fig2}
\end{figure}

In the first scenario, the underlying true state evolves according to a Markov chain with equal exit probabilities, which is the same mechanism on which our proposed algorithm is based. The true exit probability is set to $\alpha_0=0.1$, and we tune the parameter $\alpha$ with different fixed values of $\sigma$ in the Diffusion $\alpha$-HMM, linearized Diffusion $\alpha$-HMM, and ASL (where $\delta$ is set as $\alpha M$). The probability of successfully tracking the true state over 50,000 time steps is computed, and the comparison results are illustrated in Fig.~\ref{NE_fig2}. 

A horizontal comparison in Fig.~\ref{NE_fig2} reveals that although the learning accuracy is generally higher in the fully connected network, in the strongly connected network, even agent 5, which lacks distinguishing ability, can still correctly infer the true state most of the time due to information aggregation from other agents. A vertical comparison shows that as the standard deviation of noise $\sigma$ increases, the learning accuracy decreases. When comparing the three algorithms, we observe that under low noise conditions, the highest learning accuracy follows the order: Diffusion $\alpha$-HMM $>$ Linearized Diffusion $\alpha$-HMM $>$ Adaptive Social Learning. Moreover, the first two methods exhibit superior performance over a broader range of parameter values compared to ASL. Additionally, Diffusion $\alpha$-HMM consistently outperforms the linearized model, demonstrating the benefits of nonlinearity in improving learning accuracy. However, when the noise level is high, ASL achieves slightly higher maximum learning accuracy than both Diffusion $\alpha$-HMM and Linearized Diffusion $\alpha$-HMM in the fully connected network, highlighting the role of step-size scaling in mitigating noise effects.

\begin{figure}[!t]
\centerline{\includegraphics[width=\columnwidth]{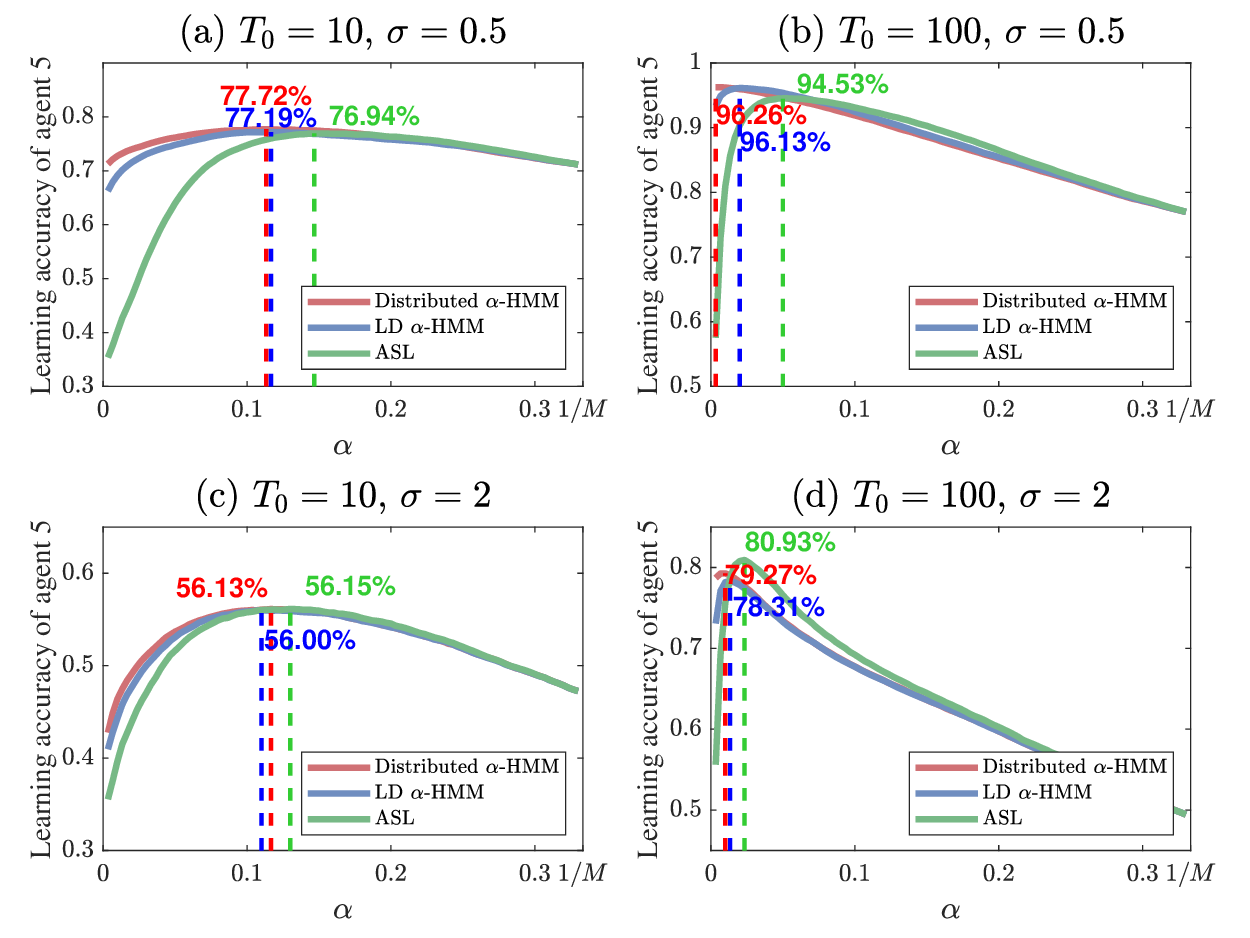}}
\caption{Comparison of the learning accuracy of agent 5, which has no distinguishing capability, under different transition intervals $T_0$ and algorithms in Scenario 2. The accuracy is evaluated as a function of $\alpha$ for different fixed values of $\sigma$. LD $\alpha$-HMM refers to the linearized Diffusion $\alpha$-HMM.}
\label{NE_fig3}
\end{figure}

In the second scenario, we assume that every $T_0$ iterations, the underlying true state is randomly selected from $\Theta$. Notably, under this assumption, the state evolution does not follow a Markov chain. In the strongly connected network, we evaluate the performance of the three algorithms under different transition intervals $T_0$ and noise standard deviations $\sigma$. From Fig.~\ref{NE_fig3}, we observe that despite the change in the state evolution mechanism, the relative performance of the three algorithms remains similar to that in Scenario 1. This further highlights the role of the nonlinear inference step in dynamic environments. Additionally, when the environment is less volatile (e.g., Fig.~\ref{NE_fig3}(b)), the performance of ASL for small $\delta$ deteriorates. This indicates that a small $\delta$ applied to observational data may hinder the learning capability of the multi-agent system.

\begin{figure}[!t]
\centerline{\includegraphics[width=0.5\columnwidth]{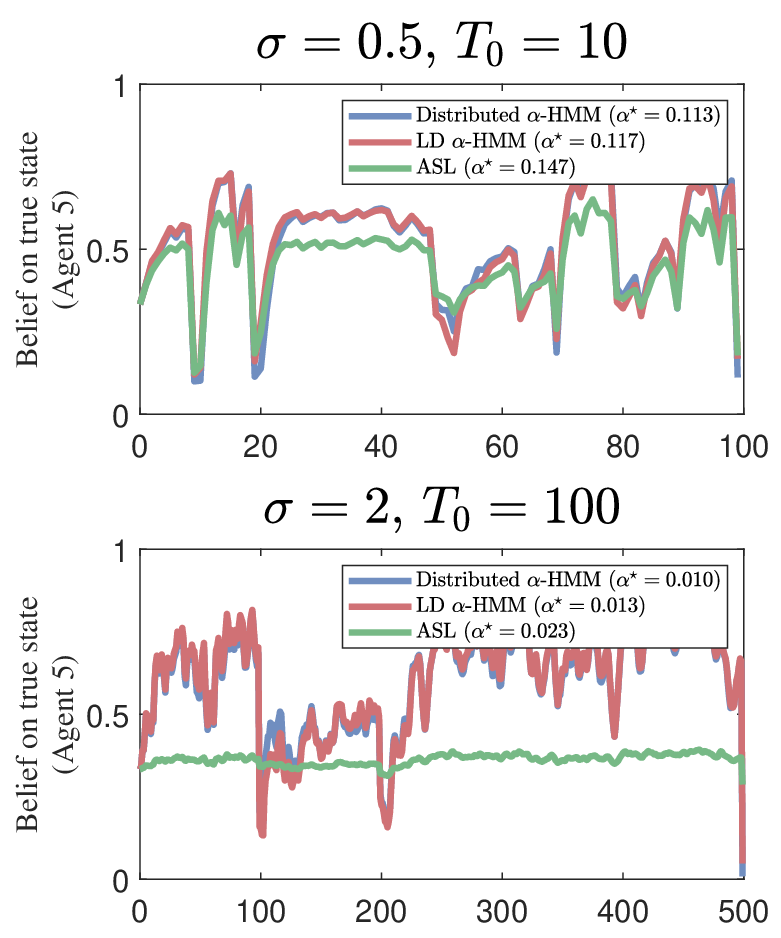}}
\caption{Comparison of agent 5's belief evolution on the true state across the three algorithms under two different settings.}
\label{NE_fig4}
\end{figure}

Fig.~\ref{NE_fig4} illustrates the evolution of agent 5's belief on the true state under Scenario 2, comparing the three algorithms with their respective optimal parameter values, $\alpha^\star$, in two different settings. From the figure, it can be observed that when the observation noise $\sigma$ is small and the environment is more dynamic (i.e., smaller $T_0$), the Diffusion $\alpha$-HMM exhibits a faster adaptation rate. Conversely, when the observation noise is large and the environment is more stable, ASL can mitigate the impact of observation noise by selecting a smaller $\alpha$ (corresponding to a smaller step-size $\delta$), resulting in a smoother belief evolution. However, in such cases, the beliefs across different states tend to be closer to each other.

%\section{Conclusion}
%In this paper, we propose an online distributed social learning strategy that enables a multi-agent network to infer an evolving environmental state through local observations and information exchange with neighbors. This strategy, named Diffusion $\alpha$-HMM, consists of two main steps. In the inference step, we assume that the environmental state evolves according to a Markov chain with equal exit probabilities. 

%By simplifying the parameterization, we derive the nonlinear dynamics governing the evolution of the log-belief ratio over time. The introduction of nonlinearity is a key distinction from previous online social learning frameworks. Theoretical analysis and numerical experiments demonstrate the connection between Diffusion $\alpha$-HMM and the Adaptive Social Learning (ASL) strategy, while also highlighting the advantages of nonlinearity in enhancing learning adaptability in dynamic environments.

\bibliographystyle{IEEEtran}
\bibliography{IEEEabrv,main}

\newpage
\appendices

\section{Proof of Lemma 1}
We first prove that for any input $x \in \mathbb{R}^{M-1}$, the function $F_m$ maps it into a closed interval. \begin{equation}\label{combine_0}
\begin{aligned}
F_m(x)&=\log\frac{(1-\alpha M)\exp(x_{m})+\alpha+\alpha\sum_{n=1}^{M-1}\exp(x_{n})}{1-\alpha M+\alpha+\alpha\sum_{n=1}^{M-1}\exp(x_{n})}\\
&=\log\frac{(1-\alpha M+\alpha)\Omega_m+\alpha+\alpha\Upsilon_m}{\alpha \Omega_m+1-\alpha M+\alpha+\alpha\Upsilon_m}\\
&\triangleq \log H(\Omega_m,\Upsilon_m),
\end{aligned}
\end{equation}
where $\Omega_m=\exp(x_m)>0$ and $\Upsilon_m=\sum_{n\neq m}\exp(x_n)>0$ are independent. With $\Upsilon_m$ fixed, differentiating with respect to $\Omega_m$ yields:
\begin{equation}
\frac{\partial H}{\partial \Omega_m}=\frac{(1-\alpha M)(\alpha\Upsilon_m+1-\alpha M+2\alpha)}{\alpha \Omega_m+1-\alpha M+\alpha+\alpha\Upsilon_m}\ge 0.
\end{equation}
Thus we derive the following bounds:
\begin{equation}\label{combine_1}
H(\Omega_m,\Upsilon_m)\le H(+\infty,\Upsilon_m)=\frac{1-\alpha M+\alpha}{\alpha},
\end{equation}
and
\begin{equation}\label{combine_2}
\begin{aligned}
H(\Omega_m,\Upsilon_m)\ge H(0,\Upsilon_m)&=\frac{\alpha+\alpha\Upsilon_m}{1-\alpha M+\alpha+\alpha\Upsilon_m}\\
&\triangleq H_0(\Upsilon_m).
\end{aligned}
\end{equation}
Differentiating $H_0(\Upsilon_m)$ with respect to $\Upsilon_m$, we obtain:
\begin{equation}
\frac{\partial H_0}{\partial \Upsilon_m}=\frac{\alpha (1-\alpha M)}{(1-\alpha M+\alpha+\alpha \Upsilon_m)^2}>0.
\end{equation}
Therefore, we further obtain:
\begin{equation}
H_0(\Upsilon_m)\le H_0(+\infty)=1,
\end{equation}
\begin{equation}\label{combine_3}
H_0(\Upsilon_m)\ge H_0(0)=\frac{\alpha}{1-\alpha M+\alpha}.
\end{equation}
Combining \eqref{combine_0}, \eqref{combine_1}, \eqref{combine_2} and \eqref{combine_3}, we conclude that for all $m=1,\cdots,M-1$:
\begin{equation}
\left|F_m(x)\right|\le \log\frac{1-\alpha M+\alpha}{\alpha}.
\end{equation}
We now introduce the following notations:
\begin{equation}\label{x_def1}
\hat{x}_{k,i}=\left[\hat{x}_{k,i}(\theta_1),\ldots,\hat{x}_{k,i}(\theta_{M-1})\right]^\top,
\end{equation}
\begin{equation}\label{x_def2}
\hat{x}_i=\left[\hat{x}_{1,i},\ldots,\hat{x}_{N,i}\right]^\top,
\end{equation}
\begin{equation}
d_k=\left[d_k(\theta_1),\ldots,d_k(\theta_{M-1})\right]^\top,
\end{equation}
\begin{equation}
d=\left[d_1,\ldots,d_N\right]^\top,
\end{equation}
\begin{equation}
F:\:\mathbb{R}^{M-1}\rightarrow \mathbb{R}^{M-1}, \quad F(x)=\left[F_1(x),\ldots,F_{M-1}(x)\right]^\top,
\end{equation}
\begin{equation}
\tilde{F}= \bigoplus _{k=1}^N F,\quad \tilde{F}(x_1,\ldots,x_N)=\left[F(x_1),\ldots,F(x_N)\right]^\top.
\end{equation}
The reference dynamics in \eqref{reference_dynamics} can be rewritten as
\begin{equation}
\hat{x}_i=\left(A^\top\otimes I_{M-1}\right)\bigoplus_{k=1}^N F(x)-\left(A^\top\otimes I_{M-1}\right)d.
\end{equation}
For $x\in\mathbb{R}^{N(M-1)}$, define the function 
\begin{equation}
G(x)\triangleq \left(A^\top\otimes I_{M-1}\right)\bigoplus_{k=1}^N F(x)-\left(A^\top\otimes I_{M-1}\right)d,
\end{equation}
we note that $G$ is a mapping from $\mathbb{R}^{N(M-1)}$ to
\begin{equation}
\left[\log\frac{\alpha}{1-\alpha M+\alpha}-\overline{d},\log\frac{1-\alpha M+\alpha}{\alpha}\right]^{N(M-1)},
\end{equation}
which forms a compact convex subset of $\mathbb{R}^{N(M-1)}$. By Brouwer's fixed-point theorem, the reference recursion \eqref{reference_dynamics} has at least one fixed point.

We now show that the fixed points of the dynamical reference system should satisfy:
\begin{equation}\label{all_less_zero}
\hat{x}_{k}^\infty(\theta_m)<0\quad \forall k=1,\ldots,N.
\end{equation}
We employ a proof by contradiction. Suppose that for some $\theta_m \in \Theta$, we have $\hat{x}_{k}^\infty(\theta_m) \ge 0$ for all $k=1,\ldots,N$. By the properties of $F_m$ in \eqref{Fm}, we establish:
\begin{itemize}
    \item If $x_{m} < 0$, 
    \begin{equation}\label{F_prop_xl0}
    x_{m} < F_m(x) < 0,
    \end{equation}
    \item If $x_{m} \geq 0$, 
    \begin{equation}\label{F_prop_xg0}
    0 \leq F_m(x) \leq x_{m}.
    \end{equation}
\end{itemize}
 For all agent $k=1,\ldots,N$, the definition of fixed point implies:
 \begin{equation}\label{fixed_point_1}
\hat{y}_k^\infty(\theta_m)=F_m(\hat{x}_k^\infty)-d_k(\theta_m),
 \end{equation}
 \begin{equation}\label{fixed_point_2}
\hat{x}_k^\infty(\theta_m)=\sum_{\ell\in\mathcal{N}_k}a_{\ell k}\hat{y}_\ell^\infty(\theta_m).
 \end{equation}
 
From \eqref{F_prop_xg0}, if $\hat{x}_{k}^\infty(\theta_m) \ge 0$ for all $k=1,\ldots,N$, then:
\begin{equation}
\hat{y}_k^\infty(\theta_m)\le \hat{x}_k^\infty(\theta_m)-d_k(\theta_m),
\end{equation}
and then we have
\begin{equation}\label{cite1}
\sum_{\ell\in\mathcal{N}_k}a_{\ell k}\hat{y}_\ell^\infty(\theta_m)\le \sum_{\ell\in\mathcal{N}_k}a_{\ell k}\hat{x}_k^\infty(\theta_m)-\sum_{\ell\in\mathcal{N}_k}a_{\ell k}d_k(\theta_m).
\end{equation}
Using the Perron vector $\pi$ of the primitive matrix $A$, we obtain:
\begin{equation}\label{contradiction1}
\begin{aligned}
&\sum_{k=1}^N\pi_k\sum_{\ell\in\mathcal{N}_k}a_{\ell k}\hat{y}_\ell^\infty(\theta_m)\\
\le &\sum_{k=1}^N\pi_k\sum_{\ell=1}^Na_{\ell k}\hat{x}_\ell^\infty(\theta_m)-\sum_{k=1}^N\pi_k\sum_{\ell=1}^Na_{\ell k}d_{\ell}(\theta_m)\\
\le & \sum_{\ell=1}^N\pi_\ell\hat{x}_\ell^\infty(\theta_m)-\sum_{\ell=1}^N\pi_{\ell}d_{\ell}(\theta_m).
\end{aligned}
\end{equation}
Noting that the elements of the Perron vector satisfy $\pi_\ell > 0$ for all $\ell=1,\ldots,N$, and according to Assumption 2, for a given $\theta_m$, there exists some $\ell \in \{1,\ldots,N\}$ such that $d_{\ell}(\theta_m) > 0$, therefore:
\begin{equation}\label{cite2}
\sum_{\ell=1}^N\pi_\ell d_{\ell}(\theta_m)>0,
\end{equation}
then Eq.~\eqref{contradiction1} contradicts with the property of the fixed-point \eqref{fixed_point_2}.

Now we establish that there exists at least one agent $k\in\left\{1,\ldots,N\right\}$ such that:
\begin{equation}
\hat{x}_{k}^\infty(\theta_m) < 0.
\end{equation}
Define the index set of agents with non-negative fixed point values as: 
\begin{equation}
\Omega^+_m=\left\{k:\hat{x}_k^\infty(\theta_m)\ge 0\right\},
\end{equation}
and let:
\begin{equation}
\overline{\Omega}^+_m=\underset{k\in\Omega^+_m}{\arg\max}\;\hat{x}_k^\infty(\theta_m).
\end{equation}
For all $k\in\overline{\Omega}^+_m$, we have:
\begin{equation}\label{contradiction_2}
\begin{aligned}
\hat{x}_{k}^\infty(\theta_m)&=\sum_{\ell\in\mathcal{N}_k}a_{\ell k}\hat{y}_\ell^\infty(\theta_m)\\
&\overset{(a)}{\le} \sum_{\ell\in\mathcal{N}_k\cap\Omega^+_m}a_{\ell k}\hat{y}_\ell^\infty(\theta_m)\\
&\le \sum_{\ell\in\mathcal{N}_k\cap\Omega^+_m}a_{\ell k}\hat{x}_\ell^\infty(\theta_m)-\sum_{\ell\in\mathcal{N}_k\cap\Omega^+_m}a_{\ell k}d_{\ell}(\theta_m),
\end{aligned}
\end{equation}
where step $(a)$ follows from the property:
\begin{equation}
\hat{y}_k^\infty(\theta_m)<-d_k(\theta_m)\le 0,\quad\text{if}\;k\notin\Omega^+_m.
\end{equation}
If there exists an $\ell \in\mathcal{N}_k$ and $\ell\notin\overline{\Omega}^+_m$, then:
\begin{equation}
\sum_{\ell\in\mathcal{N}_k\cap\Omega^+_m}a_{\ell k}\hat{x}_\ell^\infty(\theta_m)-\sum_{\ell\in\mathcal{N}_k\cap\Omega^+_m}a_{\ell k}d_{\ell}(\theta_m)<\hat{x}_k^\infty(\theta_m),
\end{equation}
which contradicts with Eq.~\eqref{contradiction_2}. Thus, for all $\ell\in\mathcal{N}_k$, we have $\ell\in\overline{\Omega}_m^+$. By iteratively applying this contradiction proof and leveraging the irreducibility of matrix $A$, we conclude that:
\begin{equation}
\Omega^+_m=\overline{\Omega}^+_m.
\end{equation}

Now for all $k\in\Omega^+_m$, we have:
\begin{equation}\label{contradiction_3_1}
\hat{x}_k^\infty(\theta_m)=\sum_{\ell\in\mathcal{N}_k}a_{\ell k}\hat{y}_\ell^\infty(\theta_m).
\end{equation}
Since matrix $A$ is irreducible, there exist $k_0\in\Omega^+_m$ and $\ell_0\notin\Omega^+_m$ such that $a_{\ell_0,k_0}>0$. Therefore:
\begin{equation}\label{contradiction_3_2}
\begin{aligned}
\sum_{\ell\in\mathcal{N}_{k_0}}a_{\ell,k_0}\hat{y}_\ell^\infty(\theta_m)&=\sum_{\ell\in\Omega^+_m}a_{\ell,k_0}\hat{y}_\ell^\infty(\theta_m)+\sum_{\ell\notin\Omega^+_m}a_{\ell,k_0}\hat{y}_\ell^\infty(\theta_m)\\
&<\sum_{\ell\in\Omega^+_m}a_{\ell,k_0}\hat{y}_\ell^\infty(\theta_m)\\
& \le \sum_{\ell\in\Omega^+_m}a_{\ell,k_0}\hat{x}^\infty_\ell(\theta_m)-\sum_{\ell\in\Omega^+_m}a_{\ell,k_0}d_{\ell}(\theta_m)\\
&\overset{(b)}{<}\hat{x}^\infty_{k_0}(\theta_m). 
\end{aligned}
\end{equation}
Step $(b)$ follows from the fact that $d_{\ell}(\theta_m)\ge 0$, and that all $\hat{x}_k^\infty(\theta_m)$ for $k\in\Omega^+_m$ has the same value, and
\begin{equation}
\sum_{\ell\in\Omega^+_m}a_{\ell,k_0}=1-\sum_{\ell\notin\Omega^+_m}a_{\ell,k_0}<1-a_{\ell_0,k_0}<1.
\end{equation}
Now we can find that Eq.~\eqref{contradiction_3_1} contradicts with Eq.~\eqref{contradiction_3_2}, leading to
\begin{equation}
\Omega^+_m=\overline{\Omega}^+_m=\emptyset.
\end{equation}
Thus, we conclude that Eq.~\eqref{all_less_zero} holds.

Finally, using \eqref{F_prop_xl0}, \eqref{fixed_point_1} and \eqref{fixed_point_2}, we establish that for all $k=1,\ldots,N$ and $m=1,\ldots,M-1$:
\begin{equation}
\hat{y}_k^\infty(\theta_m)<-d_k(\theta_m),
\end{equation}

\begin{equation}
\hat{x}_k^\infty<-\sum_{\ell\in\mathcal{N}_k}a_{\ell k}d_{\ell}(\theta_m).
\end{equation}
This completes the proof.

\section{Proof of Theorem 1}
We first claim that there exists a time instant $i_0$ such that:
\begin{equation}\label{claim}
\hat{x}_{k,i_0}(\theta_m)< 0\quad\forall k=1,\ldots,N,\;m=1,\ldots,M-1.
\end{equation}
Furthermore, for any $i>i_0$, following \eqref{F_prop_xl0}, we have:
\begin{equation}
\hat{x}_{k,i}(\theta_m)< 0\quad\forall k=1,\ldots,N,\;m=1,\ldots,M-1.
\end{equation}
To prove the claim, We utilize a linear inequality derived from the properties of the nonlinear function $F_m$. Define:
\begin{equation}
\Omega^+_{m,i}=\left\{k:\hat{x}_k^\infty(\theta_m)\ge 0\right\},
\end{equation}
If $\Omega^+_{m,i-1}$ is an empty set, then the claim is immediately proven. Otherwise, from \eqref{F_prop_xl0} and \eqref{F_prop_xg0}, we obtain:
\begin{equation}\label{linear_ineq}
\hat{x}_{k,i}(\theta_m)\le \sum_{\ell\in\mathcal{N}_k\cap\Omega_{i-1}^+}a_{\ell k}\hat{x}_{\ell,i-1}(\theta_m)-\sum_{\ell\in\mathcal{N}_k}a_{\ell k}d_{\ell}(\theta_m).
\end{equation}
Define:
\begin{equation}
U_i(\theta_m)=\underset{k=1,\ldots,N}{\max}\;\left\{\hat{x}_{k,i}(\theta_m)\right\}.
\end{equation}
From \eqref{linear_ineq}, for all $m$, we have:
\begin{equation}
U_i(\theta_m)\le U_{i-1}(\theta_m).
\end{equation}
Since we have assumed that $\Omega^+_{m,i-1}\neq \emptyset$, it follows that:
\begin{equation}
U_i(\theta_m)\ge 0,\forall i=0,1,\ldots.
\end{equation}
Thus, $U_i$ is a non-increasing and bounded sequence. By the Monotone Convergence Theorem, we conclude:
\begin{equation}\label{limit}
\lim_{i\rightarrow\infty}U_i(\theta_m)=U_m\ge 0.
\end{equation}
Now we are going to establish a contradiction. For an arbitrary $\varepsilon>0$, there exists a time instant $i_1$ such that for all $i>i_1$:
\begin{equation}
U_i(\theta_m)<U_m+\varepsilon.
\end{equation}
From the proof of Lemma 1 (Eq.~\eqref{cite1}, \eqref{contradiction1} and \eqref{cite2}), we know that there must exist a time instant $i_2>i_1$ such that $\exists\;k_0\in\left\{1,\ldots,N\right\}$ satisfying
\begin{equation}
\hat{x}_{k_0,i_2}=-c<0,
\end{equation}
otherwise the weighted sum $\sum_{k=1}^N\pi_k\hat{x}_{k,i}(\theta_m)$ would continue to decrease by $\sum_{k=1}^N\pi_kd_k(\theta_m)$, which is strictly positive.

At time instant $i_2+1$, let $\mathcal{I}^p_k$ denote the $p$-hop out-neighbors of agent $k$ and $a$ denote the smallest positive element in $A$, then for all $k\in\mathcal{I}^1_{k_0}$,
\begin{equation}
\begin{aligned}
\hat{x}_{k,i_2+1}(\theta_m)&\le \sum_{\ell\in\mathcal{N}_k\cap\Omega_{i_2}^+}a_{\ell k}\hat{x}_{\ell,i_2}(\theta_m)-\sum_{\ell\in\mathcal{N}_k}a_{\ell k}d_{\ell}(\theta_m)\\
&\le -a_{k_0,k}c+(1-a_{k_0,k})(U_m+\varepsilon)\\
&\le U_m-ac+O(\epsilon).
\end{aligned}
\end{equation}
Similarly, for all $k\in\mathcal{I}^2_{k_0}$,
\begin{equation}
\begin{aligned}
\hat{x}_{k,i_2+2}(\theta_m)&\le \sum_{\ell\in\mathcal{N}_k\cap\Omega_{i_2+1}^+}a_{\ell k}\hat{x}_{\ell,i_2+1}(\theta_m)-\sum_{\ell\in\mathcal{N}_k}a_{\ell k}d_{\ell}(\theta_m)\\
&\le U_m-a^2c+O(\varepsilon).
\end{aligned}
\end{equation}
Since the matrix $A$ is irreducible, by recursion, for all agent $k$, there exists a time instant $i_2+1\le i\le i_2+n$ such that:
\begin{equation}
\hat{x}_{k,i}(\theta_m)\le U_m-a^nc+O(\varepsilon).
\end{equation}
Due to the aperiodicity of matrix $A$, there exists at least one agent $k_1$ with a self-loop, i.e., $a_{k_1,k_1}\ge a>0$. For all $i\ge i_2+n+1$,
\begin{equation}
\begin{aligned}
\hat{x}_{k_1,i}(\theta_m)&\le \sum_{\ell\in\mathcal{N}_{k_1}\cap\Omega_{i-1}^+}a_{\ell k_1}\hat{x}_{\ell,i-1}(\theta_m)-\sum_{\ell\in\mathcal{N}_{k_1}}a_{\ell k_1}d_{\ell}(\theta_m)\\
&\le a_{k_1,k_1}\hat{x}_{k_1,i-1}(\theta_m)+(1-a_{k_1,k_1})(U_m+\varepsilon)\\
&\le a_{k_1,k_1}\left(U_m-a^nc\right)+(1-a_{k_1,k_1})U_m+O(\varepsilon)\\
&\le U_m-a^{n+1} c+O(\varepsilon).
\end{aligned}
\end{equation}
Applying the same recursion again due to the irreducibility of matrix $A$, we obtain that for all agent $k$ and for all $i\ge i_2+2n$:
\begin{equation}
\hat{x}_{k,i}(\theta_m)\le U_m-a^{2n}c+O(\varepsilon).
\end{equation}
Choosing $\varepsilon$ sufficiently small, there exists a time instant $i$ such that:
\begin{equation}
\hat{x}_{k,i}(\theta_m)\le U_m-a^{2n}c+O(\varepsilon)<U_m.
\end{equation}
This contradicts \eqref{limit}, completing the proof of claim \eqref{claim}.

We proceed by introducing the following definition:
\begin{definition}[Contractions]
Let $(X,d)$ be a metric space. A mapping $T: X\rightarrow X$ is a contraction, if there exists a constant $c$, with $0\le c< 1$, such that
\begin{equation}
d\left(T(x),T(y)\right)\le c d(x,y)
\end{equation}
for all $x,y\in X$.
\end{definition}

Next, we establish that the mapping 
\begin{equation}
F(x)=\left[F_1(x),\ldots,F_{M-1}(x)\right]^\top,
\end{equation}
where $F_m(x)$ is defined in \eqref{Fm} is a contraction in the region $\left(-\infty,0\right)^{M-1}$ with respect to the $L_\infty$-norm.

For any $x=\left[x_1,\ldots,x_{M-1}\right]^\top$ and $x'=\left[x_1',\ldots,x_{M-1}'\right]^\top$, by Lagrange's main value theorem, we have:
\begin{equation}
F(x)-F(x')=J_F(c)\left(x-x'\right),
\end{equation}
where $J_F(c)$ is the Jacobian matrix of $F$ at $c=tx+(1-t)x'$ with $t\in(0,1)$.

Using the Cauchy-Schwartz inequality:
\begin{equation}\label{main_value}
\left\|F(x) - F(x')\right\|_\infty \leq \left\|J_F(c)\right\|_\infty \left\|x - x'\right\|_\infty.
\end{equation}
The function $F_m(x)$ is given by:
\begin{equation}
F_m(x)=\log\Lambda_m(x)-\log\Gamma(x),
\end{equation}
where 
\begin{equation}
\Lambda_m(x)=(1-\alpha M)\exp(x_m)+\alpha+\alpha\sum_{n=1}^{M-1}\exp(x_n),
\end{equation}
\begin{equation}
\Gamma(x)=1-\alpha M+\alpha+\alpha\sum_{n=1}^{M-1}\exp(x_n).
\end{equation}
The partial derivative of $F_m(x)$ with respect to $x_\ell$ ($\ell\neq m$) is:
\begin{equation}
\begin{aligned}
\frac{\partial F_m}{\partial x_\ell}(x)&=\frac{\alpha\exp(x_\ell)}{\Lambda_m(x)}-\frac{\alpha\exp(x_\ell)}{\Gamma(x)}\\
&=\frac{\alpha(1-\alpha M)\exp(x_\ell)\left(1-\exp(x_m)\right)}{\Lambda_m(x)\Gamma(x)}.
\end{aligned}
\end{equation}
Since $0<\alpha<1/M$, $\Lambda_m(x)>0$ and $\Gamma(x)>0$, we have:
\begin{equation}
\frac{\partial F_m}{\partial x_\ell}(x)>0,\quad\text{if } x_m<0.
\end{equation}
Similarly, for all $m=1,\ldots, M-1$, we have:
\begin{equation}
\begin{aligned}
\frac{\partial F_m}{\partial x_m}(x)&=\frac{(1-\alpha M+\alpha)\exp(x_m)}{\Lambda_m(x)}-\frac{\alpha\exp(x_m)}{\Gamma(x)}
\\&=\frac{(1-\alpha M)\exp(x_m)\left(\alpha+\Gamma(x)-\alpha\exp(x_m)\right)}{\Lambda_m(x)\Gamma(x)}.
\end{aligned}
\end{equation}
Since $\Gamma(x)>\alpha\exp(x_m)$ for all $m$, we have:
\begin{equation}
\frac{\partial F_m}{\partial x_m}(x)>0.
\end{equation}
Summing over $\ell$, the Jacobian satisfies:
\begin{equation}\label{sum_jacob}
\begin{aligned}
\sum_{\ell=1}^{M-1}\frac{\partial F_m}{\partial x_\ell}(x)=&\frac{(1-\alpha M)\exp(x_m)+\alpha\sum_{\ell=1}^{M-1}\exp(x_\ell)}{\Lambda_m(x)}\\
&-\frac{\alpha\sum_{\ell=1}^{M-1}\exp(x_\ell)}{\Gamma(x)}\\
=&\frac{\Lambda_m(x)-\alpha}{\Lambda_m(x)}-\frac{\Gamma(x)-(1-\alpha M+\alpha)}{\Gamma(x)}\\
=&\frac{1-\alpha M+\alpha}{\Gamma(x)}-\frac{\alpha}{\Lambda_m(x)}.
\end{aligned}
\end{equation}
For $x<0$, we have $\exp(x)<1$, leading to:
\begin{equation}
\alpha<\Lambda(x)<1,
\end{equation}
\begin{equation}
1-\alpha M+\alpha<\Gamma(x)<1.
\end{equation}
Thus for all $m=1,\ldots,M-1$:
\begin{equation}\label{sum_lt}
\sum_{\ell=1}^{M-1}\frac{\partial F_m}{\partial x_\ell}(x)<1-\alpha,\quad\text{if }x_m<0.
\end{equation}
For $x,x'\in\left(-\infty,0\right)$, $c=tx+(1-t)x'\in \left(-\infty,0\right)$,
and for all $m,\ell=1,\ldots,M-1$, we have:
\begin{equation}
\left[J_F(c)\right]_{m\ell}=\frac{\partial F_m}{\partial x_\ell}(c)>0,
\end{equation}
and
\begin{equation}
\begin{aligned}
\left\|J_F(c)\right\|_\infty&=\underset{m=1,\ldots,M-1}{\max}\sum_{\ell=1}^{M-1}\left|\frac{\partial F_m}{\partial x_\ell}(c)\right|\\&=\underset{m=1,\ldots,M-1}{\max}\sum_{\ell=1}^{M-1}\frac{\partial F_m}{\partial x_\ell}(c)\\
&<1-\alpha.
\end{aligned}
\end{equation}
Following \eqref{main_value} we obtain:
\begin{equation}\label{contraction_rate}
\left\|F(x) - F(x')\right\|_\infty \leq (1-\alpha) \left\|x - x'\right\|_\infty,
\end{equation}
which completes the proof that $F(x)$ is a contraction in region $\left(-\infty,0\right)^{M-1}$ with respect to $L_\infty$-norm.

Following the notation in \eqref{x_def1} and \eqref{x_def2}, we establish the following contraction property for all $\hat{x}_{i-1} < 0$:

\begin{equation}\label{multiagent_contraction}
\begin{aligned}
&\left\|\hat{x}_{i}-\hat{x}^\infty\right\|_\infty\\
=&\left\|\left(A^\top\otimes I_{M-1}\right)\left(\bigoplus_{k=1}^NF(\hat{x}_{i-1})-\bigoplus_{k=1}^N F(\hat{x}^\infty)\right)\right\|_\infty\\
\overset{(a)}{\le} &  \left\|A^\top\otimes I_{M-1}\right\|_\infty\left\|\bigoplus_{k=1}^NF(\hat{x}_{i-1})-\bigoplus_{k=1}^N F(\hat{x}^\infty)\right\|_\infty\\
\overset{(b)}{\le} & (1-\alpha)\left\|\hat{x}_{i-1}-\hat{x}^\infty\right\|_\infty.
\end{aligned}
\end{equation}
Here, step $(a)$ follows from the Cauchy-Schwarz inequality, and step $(b)$ holds due to \eqref{contraction_rate} and the fact that $A$ is a left-stochastic matrix.

Finally, we invoke the following lemma to complete the proof:
\begin{lemma}[Contraction mapping\cite{hunter2001applied}]
If $T: X\rightarrow X$ is a contraction on a complete metric space $(X,d)$, then there is exactly one solution $x\in X$ of $T(x)=x$.
\end{lemma}

Starting from the time instant $i$ when
\begin{equation}
\hat{x}_{k,i}(\theta_m)< 0\quad\forall k=1,\ldots,N,\;m=1,\ldots,M-1,
\end{equation}
the sequence $\hat{x}_{i}$ will always remain within $\left(-\infty,0\right)^{N(M-1)}$. Therefore, by Lemma 2 and the fact that \eqref{multiagent_contraction} holds, there exists a unique fixed point $\hat{x}^\infty\in \left(-\infty,0\right)^{N(M-1)}$, such that
\begin{equation}
\lim_{i\rightarrow\infty}\hat{x}_{i}=\hat{x}^\infty.
\end{equation}

\section{Proof of Theorem 2}
Define 
\begin{equation}
\boldsymbol{\delta}_{k,i}(\theta_m)=\sum_{\ell\in\mathcal{N}_k}a_{\ell k}\log\frac{L_\ell(\boldsymbol{\xi}_{\ell,i}|\theta_m)}{L_\ell(\boldsymbol{\xi}_{\ell,i}|\theta_0)}+\sum_{\ell\in\mathcal{N}_k}a_{\ell k}d_{\ell}(\theta_m),
\end{equation}
\begin{equation}
\boldsymbol{\delta}_{k,i}=\left[\boldsymbol{\delta}_{k,i}(\theta_1),\ldots,\boldsymbol{\delta}_{k,i}(\theta_{M-1})\right]^\top,
\end{equation}
and
\begin{equation}
\boldsymbol{\delta}_{i}=\left[\boldsymbol{\delta}_{1,i},\ldots,\boldsymbol{\delta}_{N,i}\right]^\top.
\end{equation}
Then, the original and reference dynamics can be expressed as:
\begin{equation}\label{app_recursion1}
\boldsymbol{x}_{k,i}(\theta_m)=\sum_{\ell\in\mathcal{N}_k}a_{\ell k}F_m(\boldsymbol{x}_{\ell,i-1})-\sum_{\ell\in\mathcal{N}_k}a_{\ell k}d_{\ell}(\theta_m)+\boldsymbol{\delta}_{k,i}(\theta_m),
\end{equation}
where
\begin{equation}
\boldsymbol{x}_{\ell,i}=\left[\boldsymbol{x}_{\ell,i}(\theta_1),\ldots,\boldsymbol{x}_{\ell,i}(\theta_{M-1})\right]^\top,
\end{equation}
and
\begin{equation}\label{app_recursion2}
\hat{x}_{k,i}(\theta_m)=\sum_{\ell\in\mathcal{N}_k}a_{\ell k}F_m(\hat{x}_{\ell,i-1})-\sum_{\ell\in\mathcal{N}_k}a_{\ell k}d_{\ell}(\theta_m),
\end{equation}
with
\begin{equation}
\hat{x}_{k}^\infty(\theta_m)=\sum_{\ell\in\mathcal{N}_k}a_{\ell k}F_m(\hat{x}^\infty_{\ell})-\sum_{\ell\in\mathcal{N}_k}a_{\ell k}d_{\ell}(\theta_m).
\end{equation}

If the assumption in Theorem 2, i.e. Eq.~\eqref{data_quality} holds, due to the property \eqref{F_prop_xl0}, for all $k=1,\ldots,N$, $m=1,\ldots,M-1$ and $i=1,2,\ldots$,
\begin{equation}\label{noise}
\begin{aligned}
\boldsymbol{x}_{k,i}(\theta_m)=&\sum_{\ell\in\mathcal{N}_k}a_{\ell k}F_m(\boldsymbol{x}_{\ell,i-1})-\sum_{\ell\in\mathcal{N}_k}a_{\ell k}d_{\ell}(\theta_m)+\boldsymbol{\delta}_{k,i}(\theta_m)\\
<& \sum_{\ell\in\mathcal{N}_k}a_{\ell k}F_m(\boldsymbol{x}_{\ell,i-1})-\sum_{\ell\in\mathcal{N}_k}a_{\ell k}d_{\ell}(\theta_m)+C\\
<&\sum_{\ell\in\mathcal{N}_k}a_{\ell k}F_m(\boldsymbol{x}_{\ell,i-1})- \tilde{d},
\end{aligned}
\end{equation}
where 
\begin{equation}
\tilde{d}=\underset{k=1,\ldots,N}{\min}\;\underset{m=1,\ldots,M-1}{\min}\sum_{\ell\in\mathcal{N}_k}a_{\ell k}d_{\ell}(\theta_m)-C>0.
\end{equation}

Let $\overline{\boldsymbol{x}}_i(\theta_m)=\underset{k=1,\ldots,N}{\max}\boldsymbol{x}_{k,i}(\theta_m)$. According to \eqref{noise} and \eqref{F_prop_xg0}, for all $k=1,\ldots,N$ and $m=1,\ldots,M-1$, if there exist indices $k_0$ and $m_0$ such that $\boldsymbol{x}_{k_0,i-1}(\theta_{m_0}) > 0$, then the evolution of $\boldsymbol{x}_{k,i}(\theta_{m_0})$ satisfies the following inequality:
\begin{equation}\label{maximum_sequence}
\begin{aligned}
\boldsymbol{x}_{k,i}(\theta_{m_0})&<\sum_{\ell\in\mathcal{N}_k}a_{\ell k}F_{m_0}(\boldsymbol{x}_{\ell,i-1})- \tilde{d}\\
&<\sum_{\ell\in\mathcal{N}_k\cap\boldsymbol{\Omega}^+_{i-1}(\theta_{m_0})}a_{\ell k}\overline{\boldsymbol{x}}_{i-1}(\theta_{m_0})-\tilde{d}\\
&<\overline{\boldsymbol{x}}_{i-1}(\theta_{m_0})-\tilde{d},
\end{aligned}
\end{equation}
where
\begin{equation}
\boldsymbol{\Omega}^+_{i}(\theta_m)=\left\{ k:\boldsymbol{x}_{k,i}(\theta_m)>0\right\}.
\end{equation}

Since \eqref{maximum_sequence} holds for all $k=1,\ldots,N$, we obtain:

\begin{equation}
\overline{\boldsymbol{x}}_i(\theta_{m_0})<\overline{\boldsymbol{x}}_{i-1}(\theta_{m_0})-\tilde{d}.
\end{equation}

For all agents $k=1,\ldots,N$ and states $m=1,\ldots,M-1$, starting from their initial values $\boldsymbol{x}_{k,0}(\theta_m)$, after time $i_0$, where
\begin{equation}
i_0>\frac{\underset{k=1,\ldots,N}{\max}\;\underset{m=1,\ldots,M-1}{\max}\boldsymbol{x}_{k,0}(\theta_m)}{\tilde{d}},
\end{equation}

we have:

\begin{equation}
\boldsymbol{x}_{k,i_0}(\theta_m)<0,
\end{equation}

After time instant $i_0$, according to \eqref{F_prop_xl0}, for all $k=1,\ldots,N$ and $m=1,\ldots,M-1$, the update satisfies:
\begin{equation}
\boldsymbol{x}_{k,i}(\theta_{m})<\sum_{\ell\in\mathcal{N}_k}a_{\ell k}F_m(\boldsymbol{x}_{\ell,i-1})-\tilde{d}<0.
\end{equation}

Then from the recursion \eqref{app_recursion1} and \eqref{app_recursion2}, for all $i\ge i_0$ we have:
\begin{equation}
\begin{aligned}
&\mathbb{E}\left[\left.\left\|\boldsymbol{x}_{i}-\hat{x}^\infty\right\|_\infty\right|\boldsymbol{x}_{i-1}\right]\\
\overset{(a)}{\le} & \left\|A\otimes I_{M-1}\right\|_\infty\\
&\times \mathbb{E}\left[\left.\left\|\bigoplus_{k=1}^NF\left(\boldsymbol{x}_{i-1}\right)-\bigoplus _{k=1}^NF\left(\hat{x}^\infty\right)\right\|_\infty\right|\boldsymbol{x}_{i-1}\right]\\&+\mathbb{E}\left[\left\|\boldsymbol{\delta}_i\right\|_\infty\right]\\
\overset{(b)}{\le} & (1-\alpha) \mathbb{E}\left[\left.\left\|\boldsymbol{x}_{i-1}-\hat{x}^\infty\right\|_\infty\right|\boldsymbol{x}_{i-1}\right]+C.
\end{aligned}
\end{equation}
Here step $(a)$ is due to Cauchy-Schwarz inequality and triangle inequality, step $(b)$ holds due to \eqref{contraction_rate}, $A$ is a left-stochastic matrix and Assumption 4.

By taking expectation again, we obtain that for all $i\ge i_0$,
\begin{equation}\label{expect_gap}
\begin{aligned}
&\mathbb{E}\left[\left\|\boldsymbol{x}_{i}-\hat{x}^\infty\right\|_\infty\right]\\
\le& (1-\alpha)\mathbb{E}\left[\left\|\boldsymbol{x}_{i-1}-\hat{x}^\infty\right\|_\infty\right]+C\\
\le & (1-\alpha)^{i-i_0} \mathbb{E}\left[\left\|\boldsymbol{x}_{i_0}-\hat{x}^\infty\right\|_\infty\right]+\frac{C(1-(1-\alpha)^{i-i_0})}{\alpha}\\
\le & O((1-\alpha)^{i-i_0})+\frac{C}{\alpha}.
\end{aligned}
\end{equation}

The instantaneous error probability can be then computed when $i\ge i_0$ as:
\begin{equation}\label{error_probability_compute}
\begin{aligned}
&p_{i}^e\\
=&\mathbb{P}\left[\;\exists\;k=1,\ldots,N\text{ and }\theta_m \neq \theta_0, \text{ s.t. } \boldsymbol{\mu}_{k,i}(\theta_m) \geq \boldsymbol{\mu}_{k,i}(\theta_0)\right]\\
= & \mathbb{P}\left[\;\exists\;k=1,\ldots,N\text{ and }\theta_m \neq \theta_0, \text{ s.t. } \boldsymbol{x}_{k,i}(\theta_m)\ge 0\right]\\
\le & \mathbb{P}\left[\underset{k=1,\ldots,N}{\max}\:\underset{m=1,\ldots,M-1}{\max}\boldsymbol{x}_{k,i}(\theta_m)-\hat{x}_k^\infty(\theta_m)\ge -\overline{x}^\infty\right]\\
\le & \mathbb{P}\left[\underset{k=1,\ldots,N}{\max}\:\underset{m=1,\ldots,M-1}{\max}\left|\boldsymbol{x}_{k,i}(\theta_m)-\hat{x}_k^\infty(\theta_m)\right|\ge -\overline{x}^\infty\right]\\
= & \mathbb{P}\left[\left\|\boldsymbol{x}_i-\hat{x}^\infty\right\|_\infty\ge -\overline{x}^\infty\right]\\
\overset{(a)}{\le}&\frac{\mathbb{E}\left[\left\|\boldsymbol{x}_{i}-\hat{x}^\infty\right\|_\infty\right]}{-\overline{x}^\infty}\\
\overset{(b)}{\le} & \frac{O((1-\alpha)^{i-i_0})+C}{-\alpha \overline{x}^\infty}.
\end{aligned}
\end{equation}
where step $(a)$ is due to Markov's inequality and step $(b)$ follows from \eqref{expect_gap}.

Finally we have the following result:
\begin{equation}
\lim_{i\rightarrow\infty} p_i^e\le \frac{C}{-\alpha \overline{x}^\infty}.
\end{equation}
\end{document}